\documentclass[journal,twocolumn]{IEEEtran} %
\usepackage[T1]{fontenc}
\usepackage[latin9]{inputenc}
\usepackage{color}
\usepackage{url}
\usepackage{amstext}
\usepackage{graphicx}
\usepackage{xcolor}
\usepackage{amssymb}
\usepackage[unicode=true,
 bookmarks=true,bookmarksnumbered=true,bookmarksopen=true,bookmarksopenlevel=1,
 breaklinks=false,pdfborder={0 0 0},pdfborderstyle={},backref=false,colorlinks=false]
 {hyperref}

\makeatletter
 \let\oldforeign@language\foreign@language
 \DeclareRobustCommand{\foreign@language}[1]{%
   \lowercase{\oldforeign@language{#1}}}

\usepackage[caption=false,font=footnotesize]{subfig}
\usepackage{flushend}

\@ifundefined{showcaptionsetup}{}{%
 \PassOptionsToPackage{caption=false}{subfig}}
\usepackage{subfig}
\makeatother

\begin{document}

\title{}

\title{Distributed Antenna Selection for Massive MIMO using Reversing Petri
Nets}

\author{Harun Siljak,~\IEEEmembership{Member,~IEEE,} Kyriaki Psara, and~Anna Philippou\thanks{Harun Siljak is with the CONNECT Centre, Trinity College Dublin, Ireland (harun.siljak@tcd.ie). Kyriaki Psara and Anna Philippou are with the Department of Computer Science, University of Cyprus (\{kpsara01, annap\}@cs.ucy.ac.cy). This publication has emanated from research supported in part by a
research grant from Science Foundation Ireland (SFI) and is co-funded
under the European Regional Development Fund under Grant Number 13/RC/2077, European Union\textquoteright s
Horizon 2020 programme under the Marie Sk\l odowska-Curie
grant agreement No 713567, and
COST Action IC1405 on Reversible Computation - extending horizons
of computing.}}

\maketitle
\begin{abstract}
Distributed antenna selection for Distributed Massive MIMO (Multiple
Input Multiple Output) communication systems reduces computational
complexity compared to centralised approaches, and provides high fault
tolerance while retaining diversity and spatial multiplexity. We propose
a novel distributed algorithm for antenna selection and show its advantage
over existing centralised and distributed solutions. The proposed
algorithm is shown to perform well with imperfect channel state information,
and to execute a small number of simple computational operations per
node, converging fast to a steady state. We base it on Reversing Petri
Nets, a variant of Petri nets inspired by reversible computation,
capable of both forward and backward execution while obeying conservation
laws. 
\end{abstract}


\section{Introduction}

\IEEEPARstart{A}{}ntenna selection in distributed Massive MIMO (Multiple
Input Multiple Output) antenna arrays is an important optimisation
problem on a complex system comprised of a large number of simple,
similar-behaving components. It is possible to retain the advantages
of a large antenna array, including interference suppression, spatial
multiplexing and diversity \cite{ozgur2013spatial} while reducing
the number of radio frequency (RF) chains and number of antennas to
power at a time, as using all available antennas is not optimal; some
antennas fail to contribute to the service \cite{hoydis_massive_2013}.
Optimal transmit antenna selection for large antenna arrays is computationally
demanding \cite{gao2018massive}, so suboptimal approaches are pursued
for real time use. A popular one is is the greedy algorithm \cite{gharavi-alkhansari_fast_2004}
where an iterative procedure adds antennas that contribute the most
to the capacity of the set of antennas already selected. This approach
has guaranteed performance bounds for receive antenna selection, but
not for the transmit case. In a different perspective, random antenna
selection was proposed as a computationally inexpensive alternative
with satisfying results in some scenarios \cite{lee_energy_2013}.
Finally, distributed decision-making has been suggested as a natural
approach to distributed massive MIMO antenna selection \cite{siljak2018distributing},
ensuring that highly correlated spatially clustered antennas are not
prioritised in selection. 

We use the distributed paradigm of Petri nets to improve the distributed decision-making. Petri nets are a powerful mathematical and graphical notation for designing, analysing and controlling a wide range of systems. For instance, they
have been applied in higher layers of the ISO-OSI model for wireless
networks \cite{heindl2001performance}. Here we employ a variant
of Petri nets, reversing Petri nets (RPN) \cite{philippou2017reversible},
a formalism capable of reversing its evolution and, as such,
conserves information. 
In this work 
we exploit its conservation of tokens. Our use of RPN is motivated by: (1) the ability of Petri nets to formalise handling of a complex system: a large number of simple entities acting asynchronously, (2) the reversibility of it, inherently allowing backtracking, periodic behaviour, fault recovery and conservation of information in the system.

In this letter we propose a fast, environment-aware, asynchronous,
distributed antenna selection algorithm maximising sum-capacity within
an RPN scheme of a distributed array. A small number of simple computational
tasks is performed, so the algorithm is simpler and faster than the
state of the art in both centralised and distributed antenna selection
algorithms.

\section{RPNs and Transmit Antenna Selection}

\subsection{The Optimisation Problem}

We consider the scenario of downlink (transmit) antenna selection
at a distributed massive MIMO base station with $N_{T}$ antennas,
with the task of selecting a subset of antennas of size $N_{TS}$. In the cell
there are $N_{R}$ single antenna users and we aim at maximising the
sum-capacity

\begin{equation}
\mathcal{C}=\max_{\mathbf{P},\mathbf{H_{c}}}\log_{2}\det\left(\mathbf{I}+\rho\frac{N_{R}}{N_{TS}}\mathbf{H_{c}}\mathbf{P}\mathbf{H_{c}}^{H}\right)\label{eq:capac}
\end{equation}
where $\rho$ is the signal to noise ratio (SNR), $\mathbf{I}$ a
$N_{TS}\times N_{TS}$ identity matrix, and $\mathbf{P}$ a diagonal $N_{R}\times N_{R}$
power distribution matrix. $\mathbf{H_{c}}$ is the $N_{TS}\times N_{R}$
channel submatrix for a selected subset of antennas from the $N_{T}\times N_{R}$
channel matrix $\mathbf{H}$. This is the approach taken in \cite{gao_massive_2015}, and we follow it here, defining the antenna selection as a problem of maximal sum-capacity for a massive MIMO system formed by the selected number of antennas, calculated for optimal (dirty paper coding) pre-coding. Following the example of \cite{NGao} we scale the power by the number of selected antennas, using the array gain to reduce transmit power per antenna, instead of increasing receive SNR.

The receiver selection problem is different from (\ref{eq:capac}) as
it does not feature scaling by the number of transmit antennas. As
such, receiver selection problem can be solved using greedy algorithms with a guaranteed (suboptimal)
performance bound by the previously described greedy algorithm. The
transmitter antenna selection is not submodular \cite{vaze_sub-modularity_2012};
adding a new antenna to the already selected set of
antennas can in fact decrease channel capacity if the contribution is under the average. Greedy algorithms
do not have a performance guarantee in this case.

The optimisation problem in (\ref{eq:capac}) has two variables, the
subset of selected antennas and the optimal power distribution over
them. Following the practice from \cite{gao_massive_2015,siljak2018distributing},
we start from the assumption of $\mathbf{P}$ having all diagonal
elements equal to $1/N_{R}$ (their sum is unity, making the total
power equal to $\rho N_{R}/N_{TS}$), and after the antenna selection
optimise $\mathbf{P}$ by water filling for zero forcing. This was suggested in \cite{gao_massive_2015} as a practicality measure; hence the results in this paper are presented after linear beamforming, and with coherent data transmission in mind. We use the Ilmprop channel model \cite{galdo2005geometry} which, among other features, accounts for spatial correlation, pathloss and shadow fading \cite{GDG}. In this consideration, the channel state information (CSI) is assumed to be perfect and the matrix $\mathbf{H}$ known in its entirety, but we show that the algorithm is robust under uncertainties and errors in $\mathbf{H}$.

\subsection{The RPN Algorithm}
We provide a general RPN model whose behaviour simulates the runs of the proposed antenna selection algorithm. The graphical representation of the RPN is lucid enough for understanding and explaining the complex structures of the distributed algorithm and it also provides a formal semantics where verification techniques can be applied. The RPN framework is independent of the array structure (or more general, topology of a network) as it does not depend on the number of antennas or on the way the antennas are connected to each other. This suggests its generalistic nature as a framework for resource allocation in wireless communications.

\begin{figure}[tbh]
\centering
\subfloat[]{
\includegraphics[width=0.5\columnwidth]{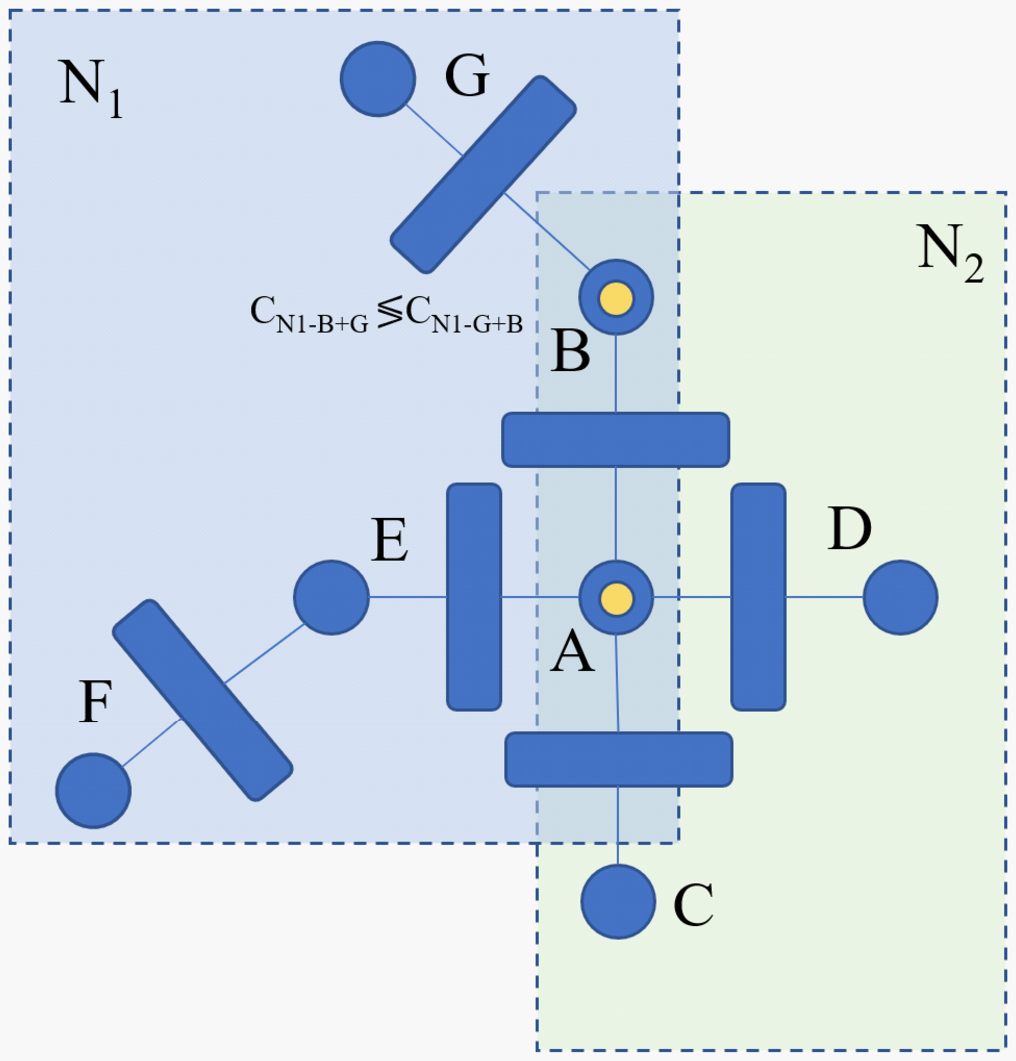}
}

\subfloat[]{
\includegraphics[width=0.8\columnwidth]{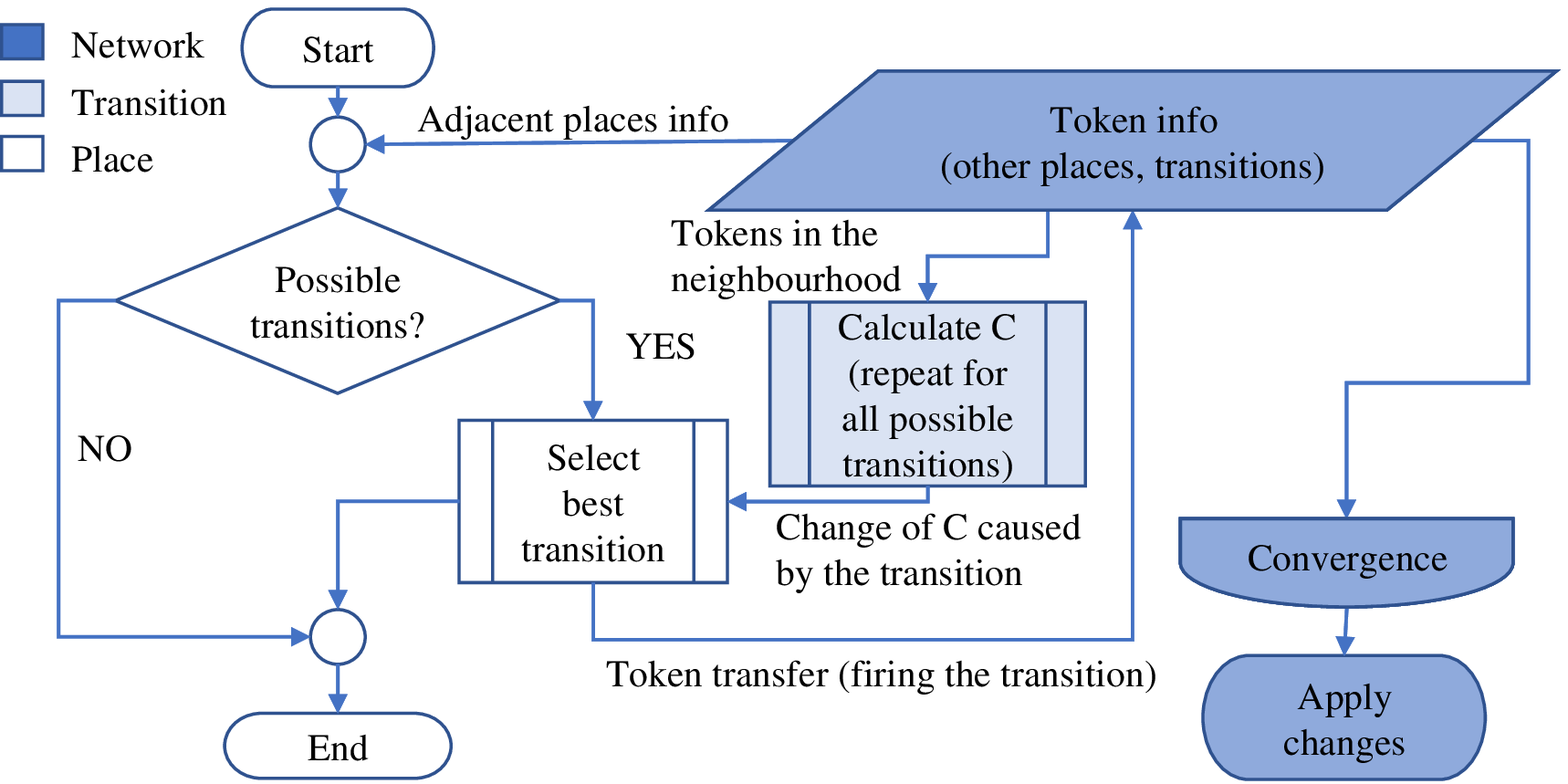}
}
\caption{(a) Exemplary Petri net for antenna selection, (b) the flowchart}
\label{explain}
\end{figure}

The algorithm we propose is illustrated in Fig. \ref{explain}, with
more information about the formal model in \cite{psara2019modelling}.
The antennas are represented with \emph{places} (circles A-G), and
the \emph{token} (bright circle) in some of the places indicates that
the respective antenna is currently
switched on. The places are divided into overlapping sets we call \emph{neighbourhoods}
($N_{1}$ and $N_{2}$ in Fig. \ref{explain}(a)) such that each two adjacent
places belong to (at least one) common neighbourhood. \emph{Transitions}, depicted as bars
between places, allow tokens to move and they operate as follows:
\begin{enumerate}
\item A transition is possible if there is a token in exactly one of the two
places (e.g. B and G in Fig. \ref{explain}) it connects. Otherwise
(e.g. A and B, or E and F) it is not possible.
\item An enabled transition will occur if the sum capacity (\ref{eq:capac})
calculated for all antennas with a token in the neighbourhood shared
by the two places (for B and G, that is neighbourhood $N_{1}$) is
less than the sum capacity calculated for the same neighbourhood,
but with the token moved to the empty place (for B-G transition,
$\mathcal{C}_{AB}<\mathcal{C}_{AG}$ ). Otherwise, it does
not occur.
\item In case of several possible transitions from one place (e.g. A-E, A-D, A-C) the one with the greatest sum-capacity difference is given priority (prioritisation is implemented in the transition condition function).
\item There is no designated order in transition execution, and transitions are
performed until a stable state is reached.
\end{enumerate}

This asynchronous scheme (represented as a flowchart in Fig. \ref{explain}(b)) requires few (in our trials, up to five for 64 Tx/16 Rx case) passes
through the network to converge to a stable state, starting
from a random selection of $n$ antennas ($n$ tokens in random
places). It conserves the number of tokens and keeps at most one token
per place, hence the resulting state will capture the set of $n$ selected antennas.

Fast convergence suggests that, for a
small number of users, the places corresponding to antennas close to particular users might not always receive tokens if we start from uniformly random initial conditions.
Thus, if the goal is to serve a relatively small
number of users ($\lesssim \sqrt{N_T}$), running several RPNs in parallel (in our experiments as few as five
is enough) and taking the best result among them is an option. For
larger number of users ($\gtrsim \sqrt{N_T}$), our results suggest that one RPN is enough.
Once the RPN converges, the  state of antennas
is changed and the antennas with tokens are turned on for the duration
of the coherence interval. At the next update of the channel state information, 
the algorithm resumes its operation.

\section{Other methods: centralised and distributed}

The computational footprint of the described algorithm is very small:
two small matrix multiplications and determinant calculations are
performed at a node which contains a token in a small number of iterations.
As such, this algorithm is significantly faster and computationally
less demanding than the centralised greedy approach which is a low-complexity
representative of global optimisation algorithms in antenna selection
\cite{gao2018massive}. In \cite{siljak2018distributing} it was shown
that another distributed algorithm, which we will call NN (Nearest
Neighbours) 
has smaller computational complexity than
the greedy algorithm, so we proceed with a comparison of the RPN and NN
approaches. In brief, every antenna element in NN calculates the sum-capacity for the currently selected antennas among its nearest neighbours and checks whether this quantity increases or decreases by including itself into the selected set.

It has been shown that the worst case complexity of NN for $N_{T}$ antennas is $\mathcal{O}(N_{T}^{\omega})$,
where $\omega$, $2<\omega<3$
is the exponent in the employed matrix multiplication algorithm complexity.
Similarly to RPN, NN performs two relatively small matrix multiplications
and determinant calculations per iteration at a node, but there are
important differences:
\begin{enumerate}
\item NN performs calculations at each node in each iteration. RPN does it
only at nodes that contain tokens.
\item NN requires large neighbourhoods $\approx N_{T}$ (i.e. large matrix
multiplications) to select small number of antennas, while RPN performs
calculations on a small neighbourhood for any number of antennas to
be selected (in our experiments, the value of $\lfloor \sqrt{N_{T}}\rfloor$ or less
is enough for the neighbourhood size).
\item NN does not converge, so the number of iterations has to be large
to pass through many different states and pick the best. RPN converges
fast (in our experiments, always under 5 iterations).
\end{enumerate}
The worst case complexity of the RPN-based approach operating on the neighbourhood size of $\lfloor N_{T}^{1/a}\rfloor$, $a>1$, is $\mathcal{O}(N_{T}^{\omega/a})$. As neighbourhood of $\lfloor \sqrt{N_{T}}\rfloor$ antennas was enough in our practical considerations, our implementation had the complexity $\mathcal{O}(N_{T}^{\omega/2})$. At the same time, the constant factor
multiplying the complexity is reduced because of fewer computing nodes
(only those with tokens) and fewer iterations (50 vs. 5). While the
this does not affect the asymptotic complexity, it
affects the number of floating point operations (flops) performed; in the next section we examine this effect.

Another advantage of RPN over NN is the ability to select how many
antennas to use, by simply choosing how many tokens to start the process
with. In NN, the number of antennas for which the system reaches the
best performance is an emergent property and as such cannot be controlled.
Furthermore, in \cite{siljak2018distributing} it was seen that the
NN algorithm requires more than $N_{R}$ antennas to serve $N_{R}$
users; RPN gives meaningful results and good performance even at $N_{R}$
antennas selected.

\section{Results and Discussion}

\begin{figure}[tbh]
\begin{centering}
\subfloat[Distribution of antennas in space]{
\includegraphics[width=.7\columnwidth]{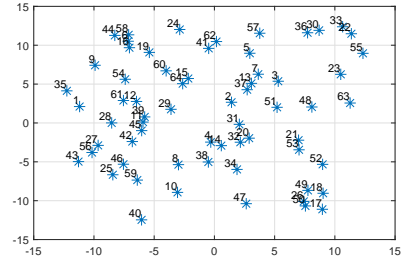}
}

\par\end{centering}
\begin{centering}
\subfloat[Position of antennas in RPN topology]{
\includegraphics[width=0.7\columnwidth]{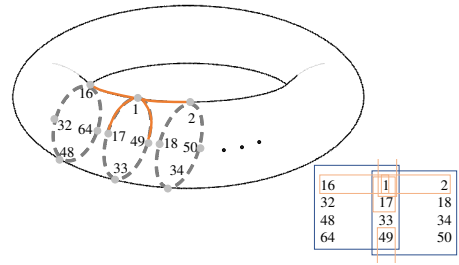}
}
\par\end{centering}
\caption{Mapping of antennas onto the toroid, with the connections and neighbourhoods
used}
\label{toro}
\end{figure}

The algorithm was tested in the same conditions as the NN algorithm
in \cite{siljak2018distributing}, using the raytracing Matlab tool Ilmprop
\cite{galdo2005geometry} with 64 distributed transmitters (Fig. \ref{toro}(a)) in the
area which included 75 scatterers and one large
obstacle. The number of users varied from 4
to 16; users, antennas, and scattering clusters were distributed uniformly, and we used SNR $\rho=-5\ \text{dB}$, 2.6
GHz carrier frequency, 20 MHz bandwidth and omnidirectional antennas
for transmitters and receivers.
In all computations, the matrix $\mathbf{H}$ was normalised to unit average energy over
all antennas, users and subcarriers \cite{gao_massive_2015}. Fig. \ref{toro}(b) shows the mapping
of the antennas into the RPN topology: they are arranged in an $4\times16$
array which is then folded into a toroid, so that the edges of the
array continuously connect to the opposite edges. In this arrangement,
antenna 1 is direct neighbour of antennas 2, 16, 17 and 49. We establish
links between immediate Von Neumann (top, down, left, right) neighbours
(allow exchange of tokens between them) and set up two overlapping
8-antenna neighbourhoods: in the example of antenna 1, transitions
to antennas 16 and 17 are governed by the neighbourhood \{16, 32,
48, 64, 1, 17, 33, 49\} while the transitions to antennas 2 and 49
are governed by \{1, 17, 33, 49, 2, 18, 34, 50\}, with the same pattern
for other antennas (left and down, left neighbourhood,
right and up, right neighbourhood). 

\begin{figure}[tbh]
\begin{centering}
\includegraphics[width=.85\columnwidth]{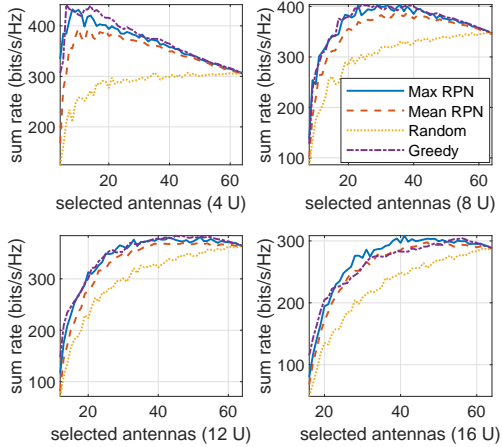}
\par\end{centering}
\caption{Achieved sum rates for 4-16 users using the proposed algorithm vs.
random and centralised greedy selection}
\label{rez}
\end{figure}

The results of the experiments for 4, 8, 12 and 16 users are shown
in Fig. \ref{rez}, comparing greedy and random selection with two
variants of our RPN approach: one as the average of five concurrently
running RPNs with random initial condition, another as the performance
of the best RPN out of those five. They demonstrate comparable performance
of the proposed algorithm to the sum rates obtained through centralised
greedy selection. Furthermore, it indicates that for a larger number
of users our proposed algorithm outperforms centralised antenna selection
represented by the greedy algorithm, while for a small number of users the
centralised algorithm performs marginally better. This is because
some regions of the distributed base station may be rightfully favoured
by the centralised algorithm in the small user pool, and yet contain
just a few tokens in our distributed algorithm; in the large user
pool, all regions of the distributed base station contribute to the
service. This in practice means that a single RPN suffices for networks
with a relatively large expected number of users. The case of many users is the one we are trying to solve, with the idea of many antennas serving many users in Massive MIMO. The need for a strategic antenna selection in all observed scenarios is demonstrated by the results of significantly under-performing random selection.

\begin{figure}[tb]
\begin{centering}
\includegraphics[width=0.85\columnwidth]{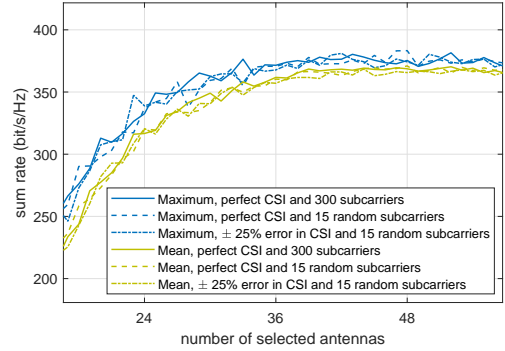}
\par\end{centering}
\caption{The effects of imperfect CSI and random selection of subcarriers}
\label{fcsi}
\end{figure}

\begin{figure}[tb]
\begin{centering}
\includegraphics[width=0.85\columnwidth]{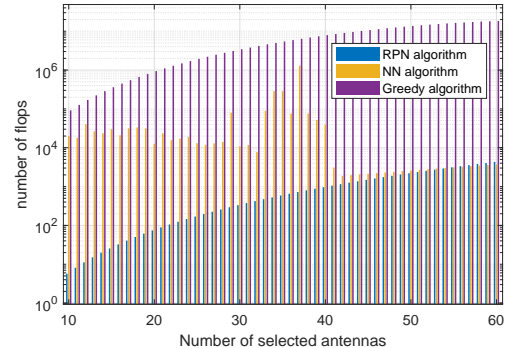}
\par\end{centering}
\caption{Comparison of the computational load}
\label{flops}
\end{figure}

In \cite{siljak2018distributing}, its has been shown that the distributed
algorithms such as NN are resistant to errors in CSI and that they
perform well even with just a (randomly selected) subset of subcarriers
used for optimisation. We performed the same test for the RPN solution,
and the result is shown in Fig. \ref{fcsi} in the case of 12 users.
In the light of the computational complexity reduction discussed in
the previous section, Fig. \ref{flops} shows the advantage of our
proposed algorithm in the number of performed calculations in the
case of choosing a number of antennas from 64 distributed antennas.

\section{Conclusions}

We have presented a novel distributed antenna selection algorithm,
improving the shortcomings of the existing distributed solutions and
additionally reducing computational complexity. Our application of
RPN is a pioneering one, and we aim to expand the RPN approach to other
resource management problems in wireless communications, drawing benefits
from both conservation properties of RPN and the ability to run the
networks, or their parts, in reverse direction to recover from faults
and handle inherently reversible communication phenomena (e.g. receiver/transmitter
duality). The ability of the RPN solution to act asynchronously and
converge fast with minimal computational burden enables real time
application of the algorithm even in high mobility scenarios. In future
work, we will investigate the ways of translating the physical topology
of the antenna array into the Petri net topology and propose solutions
for new use cases brought by 5G.

\begin{thebibliography}{10}
\providecommand{\url}[1]{#1}
\csname url@samestyle\endcsname
\providecommand{\newblock}{\relax}
\providecommand{\bibinfo}[2]{#2}
\providecommand{\BIBentrySTDinterwordspacing}{\spaceskip=0pt\relax}
\providecommand{\BIBentryALTinterwordstretchfactor}{4}
\providecommand{\BIBentryALTinterwordspacing}{\spaceskip=\fontdimen2\font plus
\BIBentryALTinterwordstretchfactor\fontdimen3\font minus
  \fontdimen4\font\relax}
\providecommand{\BIBforeignlanguage}[2]{{%
\expandafter\ifx\csname l@#1\endcsname\relax
\typeout{** WARNING: IEEEtran.bst: No hyphenation pattern has been}%
\typeout{** loaded for the language `#1'. Using the pattern for}%
\typeout{** the default language instead.}%
\else
\language=\csname l@#1\endcsname
\fi
#2}}
\providecommand{\BIBdecl}{\relax}
\BIBdecl

\bibitem{ozgur2013spatial}
A.~Ozgur, O.~L{\'e}v{\^e}que, and D.~Tse, ``Spatial degrees of freedom of large
  distributed {MIMO} systems and wireless ad hoc networks,'' \emph{IEEE Journal
  on Selected Areas in Communications}, vol.~31, no.~2, pp. 202--214, 2013.

\bibitem{hoydis_massive_2013}
J.~Hoydis, S.~t. Brink, and M.~Debbah, ``Massive {MIMO} in the {UL}/{DL} of
  {Cellular} {Networks}: {How} {Many} {Antennas} {Do} {We} {Need}?'' \emph{IEEE
  Journal on Selected Areas in Communications}, vol.~31, no.~2, pp. 160--171,
  2013.

\bibitem{gao2018massive}
Y.~Gao, H.~Vinck, and T.~Kaiser, ``Massive {MIMO} antenna selection: Switching
  architectures, capacity bounds, and optimal antenna selection algorithms,''
  \emph{IEEE Transactions on Signal Processing}, vol.~66, no.~5, pp.
  1346--1360, 2018.

\bibitem{gharavi-alkhansari_fast_2004}
M.~Gharavi-Alkhansari and A.~B. Gershman, ``Fast antenna subset selection in
  {MIMO} systems,'' \emph{IEEE Transactions on Signal Processing}, vol.~52,
  no.~2, pp. 339--347, 2004.

\bibitem{lee_energy_2013}
B.~M. Lee, J.~Choi, J.~Bang, and B.~C. Kang, ``An energy efficient antenna
  selection for large scale green {MIMO} systems,'' in \emph{2013 {IEEE}
  {International} {Symposium} on {Circuits} and {Systems} ({ISCAS} 2013)},
  2013, pp. 950--953.

\bibitem{siljak2018distributing}
H.~Siljak, I.~Macaluso, and N.~Marchetti, ``Distributing complexity: A new
  approach to antenna selection for distributed massive {MIMO},'' \emph{IEEE
  Wireless Communications Letters}, vol.~7, no.~6, pp. 902--905, 2018.

\bibitem{heindl2001performance}
A.~Heindl and R.~German, ``Performance modeling of {IEEE} 802.11 wireless
  {LAN}s with stochastic {P}etri nets,'' \emph{Performance Evaluation},
  vol.~44, no. 1-4, pp. 139--164, 2001.

\bibitem{philippou2017reversible}
A.~Philippou and K.~Psara, ``Reversible computation in {P}etri nets,'' in
  \emph{10th International Conference on Reversible Computation}.\hskip 1em
  plus 0.5em minus 0.4em\relax Springer, 2018, pp. 84--101.

\bibitem{gao_massive_2015}
X.~Gao, O.~Edfors, F.~Tufvesson, and E.~G. Larsson, ``Massive {MIMO} in {Real}
  {Propagation} {Environments}: {Do} {All} {Antennas} {Contribute} {Equally}?''
  \emph{IEEE Transactions on Communications}, vol.~63, no.~11, pp. 3917--3928,
  2015.

\bibitem{NGao}
X.~Gao, O.~Edfors, J.~Liu, and F.~Tufvesson,
  ``\BIBforeignlanguage{English}{Antenna selection in measured massive mimo
  channels using convex optimization},'' in
  \emph{\BIBforeignlanguage{English}{IEEE GLOBECOM Workshops, 2013}}, 2013.

\bibitem{vaze_sub-modularity_2012}
R.~Vaze and H.~Ganapathy, ``Sub-{Modularity} and {Antenna} {Selection} in
  {MIMO} {Systems},'' \emph{IEEE Communications Letters}, vol.~16, no.~9, pp.
  1446--1449, 2012.

\bibitem{galdo2005geometry}
G.~D. Galdo, M.~Haardt, and C.~Schneider, ``Geometry-based channel modelling of
  {MIMO} channels in comparison with channel sounder measurements,''
  \emph{Advances in Radio Science}, vol.~2, no.~BC, pp. 117--126, 2005.

\bibitem{GDG}
G.~Del~Galdo, ``Geometry-based channel modeling for multi-user mimo systems and
  applications,'' Ph.D. dissertation, Ilmenau University of Technology, Jul
  2008.

\bibitem{psara2019modelling}
A.~Philippou, K.~Psara, and H.~Siljak, ``Controlling Reversibility in Reversing {P}etri Nets with Application to Wireless Communications,'' in \emph{11th International
  Conference on Reversible Computation}, 2019.

\end{thebibliography}

\end{document}